\documentclass[12pt]{article}
\def\be{\begin{equation}} 
\def\ee{\end{equation}}   

\usepackage{graphicx,amsmath,amssymb}
\usepackage{multirow}
\usepackage{subfigure}

\usepackage{amsmath}
\usepackage{amssymb}
\usepackage{cite}

\begin{document}
\titlepage
\begin{flushright}
IPPP/17/16 \\
\today \\
\end{flushright}

\vspace*{0.5cm}
\begin{center}
{\Large \bf  $\chi_c$ decays and the gluon component of the $\eta'$, $\eta$ mesons}\\

\vspace*{1cm}

L.A. Harland-Lang$^a$, V.A.Khoze$^{b,c}$,  M.G. Ryskin$^b$ and A.G. Shuvaev$^b$

\vspace*{0.5cm}
$^a$ Department of Physics and Astronomy, University College London, WC1E 6BT, UK\\
$^b$
 Petersburg Nuclear Physics Institute, NRC Kurchatov Institute, 
Gatchina, St.~Petersburg, 188300, Russia \\
$^c$ Institute for Particle Physics Phenomenology, Durham University, Durham, DH1 3LE, UK\\ 

\end{center}

\begin{abstract}
\noindent The large mass of the $\eta'$ meson indicates that a sizeable gluon component is present in the meson wave function. However, the $\chi_{c0}$ and $\chi_{c2}$ decays to $\eta'$ mesons, which proceed via a purely gluonic intermediate state and we would therefore na\"{i}vely expect to be enhanced by such a component, are in fact relatively suppressed. We argue that this apparent contradiction may be resolved by a proper treatment of interference effects in the decay. In particular, by accounting for the destructive interference between the quark and gluon components of the $\eta'$ distribution function, in combination with a model for strange quark mass effects, we demonstrate that the observed $\chi_{c(0,2)}\to\eta(')\eta(')$ branching ratios can be reproduced for a reasonable gluon component of the $\eta'$, $\eta$ mesons.
\end{abstract}
\vspace*{0.5cm}

\section{Introduction}

The long--standing issue concerning the size of the gluon content of the $\eta'$ and $\eta$ mesons still remains unsettled, see~\cite{Kroll:2013iwa,DiDonato:2011kr,
Harland-Lang:2013ncy,Agaev:2014wna,Agaev:2015faa} for recent discussion. This question is intimately related
to important issues of non--perturbative physics, such as vacuum topology and the U(1) anomaly~\cite{'tHooft:1976snw,Crewther:1977ce,Veneziano:1979ec,Weinberg:1975ui,Glashow:1970rp} (see e.g.~\cite{Feldmann:1999uf,Shore:2007yn,Bass:2014jla} for reviews). As flavour symmetry is broken, the flavour--singlet and octet quark basis states are not degenerate in mass, and a larger $\eta$ mass than the $\eta'$ would naively be expected. However, the axial anomaly only contributes to the singlet mass, resulting in a larger $\eta'$ meson mass compared to other pseudoscalar states. A full understanding of this issue currently presents an interesting challenge for the lattice QCD, see for example~\cite{Gregory:2011sg,Fukaya:2015ara}. A variety of processes have been proposed in order to constrain the size of the two-gluon Fock state in the $\eta$ and $\eta'$ mesons. Examples of particular interest are non-leptonic exclusive isosinglet $B$-decays~\cite{Blechman:2004vc} and central exclusive production (CEP) of $\eta$, $\eta'$ pairs in proton collision~\cite{Harland-Lang:2013ncy}, as in these processes the gluon production amplitude enters at leading order. Other possibilities are discussed in for example~\cite{Escribano:2007cd,Escribano:2008rq,Mathieu:2009sg,Alte:2015dpo}.

In~\cite{Harland-Lang:2013ncy} (see also \cite{Kroll:2013iwa,Kroll:2002nt}) a puzzling issue involving the decays of  C--even charmonium $\chi_{c(0,2)}$ to $\eta$ and $\eta'$ mesons was discussed. In particular, after accounting for trivial phase space effects no enhancement with respect to the pion channels is observed experimentally~\cite{Olive:2016xmw}. As such decays proceed via a purely gluonic intermediate state this could indicate that the two-gluon component is smaller than we might generally expect, see for example~\cite{Ochs:2013gi}. Moreover, the branching ratios for the $\chi_c$ decays to $\eta\eta$ mesons are in fact observed to be larger than for the corresponding decays to $\eta'\eta'$ mesons. This is again somewhat surprising; up to limited mixing effects the $\eta'$ meson is mainly a flavour singlet state, with as mentioned above a sizeable gluonic admixture required in order to explain the larger $\eta'$ mass (the solution to the so--called $U(1)$ problem~\cite{Weinberg:1975ui,Glashow:1970rp}). Thus we would more naturally expect the $\chi_c$ decays to $\eta'\eta'$ mesons to be enhanced in comparison to $\eta\eta$, where in the latter case only a relatively small gluon component is allowed by flavour mixing. 

One possible solution to these apparent tensions, first suggested in~\cite{Harland-Lang:2013ncy}, is that destructive interference between the quark and gluon components of the pseudoscalar mesons may suppress the $\chi_{c(0,2)}$ to $\eta$ and $\eta'$ branching ratios below na\"{i}ve expectations, in particular in the $\eta'\eta'$ case. In this paper we will apply the `hard exclusive' approach developed in~\cite{Chernyak:1977as,Lepage:1979za,Lepage:1979zb,Efremov:1979qk,Lepage:1980fj} and the leading order $\chi_{c(0,2)}\to\eta(')\eta(')$ decay amplitudes calculated in~\cite{Baier:1985wv}. Using this approach, we will show that the contribution from purely gluon and purely quark final states interfere destructively, naturally leading to a suppression in the $\eta'\eta'$ mode. We will in addition demonstrate that for an appropriate choice of the gluon component of the $\eta'$ meson, the observed branching ratios of the $\chi_{c(0,2)}$ to $\eta$ and $\eta'$ mesons (or result below the upper limits where the corresponding observations are currently lacking) can be reproduced to good accuracy, resolving the apparent contradictions described above. Our numerical result is found to be consistent with that of the study of~\cite{Kroll:2013iwa} where the gluon component of the $\eta(')$ meson is extracted from an analysis of the meson transition form factors $F_{\eta(')\gamma}(Q^2)$.

The outline of this paper is as follow. In Section~\ref{sec:calc} we describe of the `hard exclusive' formalism we use. In Section~\ref{sec:amp} we give explicit expressions for the corresponding $\chi_c\to\eta(')\eta(')$ amplitudes. In Section~\ref{sec:num} we present numerical results for the $\chi_c$ branching ratios to $\eta(')$ and $\pi$ mesons, including and excluding suppression due to the strange quark mass, and comparing to experimental measurements. Finally, in Section~\ref{sec:conc} we conclude and discuss the outlook for further constraining the gluon component of the $\eta(')$ mesons.

\section{Calculation details}\label{sec:calc}

The leading order contribution to the $\chi_{c(0,2)} \to\eta(')\eta(')$ process can be calculated using the formalism described in~\cite{Chernyak:1977as,Lepage:1979za,Lepage:1979zb,Efremov:1979qk,Lepage:1980fj}, and is written in terms of the parton--level $\chi_c \to q\overline{q}q\overline{q}$, $q\overline{q}gg$, $4g$ amplitudes and the distribution amplitudes $\phi$ for the corresponding $q\overline{q}$ and $gg$ components of the $\eta(')$ mesons, see Fig.~\ref{feyn} for representative Feynman diagrams. In particular we have
\be
\mathcal{M}(\chi_c\to M_1 M_2)=\mathcal{M}_{qq}+\mathcal{M}_{qg}+\mathcal{M}_{gg}\;,
\ee
where
\be \label{mab}
\mathcal{M}_{ab}=\int_{0}^{1} \,{\rm d}x \,{\rm d}y\, \phi_{M_1}^a(x)\phi_{M_2}^b(y) T_{ab}(x,y)\;.
\ee
Here $\phi_{M}^a$ is the $a=q(g)$ distribution amplitude for the $q\overline{q}$ ($gg$) component of the meson $M$. Each $q\overline{q}$ and $gg$ pair is collinear and has the appropriate colour, spin, and flavour content projected out to form the parent meson; $x,y$ are the momentum fractions of the parent mesons carried by the quark or gluons. The amplitude $T_{ab}$ corresponds to the appropriate $\chi_c \to q\overline{q} q\overline{q}$, $q\overline{q}gg$ or $4g$ transition.

The meson distribution amplitude depends on the (non--perturbative) details of hadronic binding and cannot be predicted in perturbation theory. However, the overall normalization of the $q\overline{q}$ distribution amplitude can be set by the meson decay constant $f_q^M$ via~\cite{Brodsky:1981rp}
\begin{equation}\label{wnorm}
\int_{0}^{1}\, {\rm d}x\,\phi_M(x)=\frac{f_q^M}{2\sqrt{3}}\;.
\end{equation}
More precisely, the shape of the distribution amplitude $\phi(x,\mu_F)$ in fact depends on the factorization scale $\mu_F$, which should as usual be taken to be of the order of the characteristic hard scale of the process under consideration. It was shown in~\cite{Lepage:1980fj} that for very large $\mu_F^2$ the $q\overline{q}$ meson distribution amplitude evolves towards the asymptotic form
\begin{equation}\label{asym}
\phi_M(x,\mu_F^2)\underset{\mu_F^2\to \infty}{\to} \,\sqrt{3} f_q^M\, x(1-x)\;,
\end{equation}
where $f_q^M$ is the meson decay constant. However, at the appropriate scale $\mu_F \sim m_c$ for the $\chi_c$ decay we are far from this asymptotic region, and an alternative choice that we will make use of later on is given by~\cite{Chernyak:1981zz}
\begin{equation}\label{CZ}
\phi^{{\rm CZ}}_{q,M}(x,\mu_F^2=\mu_0^2)=5\sqrt{3}f_M\, x(1-x)(2x-1)^2\;,
\end{equation}
where the starting scale is roughly $\mu_0\approx 1$ GeV. 

More precisely, the $q\overline{q}$ flavour--singlet and $gg$ distribution amplitudes can be expanded in terms of the Gegenbauer polynomials $C_n$~\cite{Terentev:1980qu,Shifman:1980dk,Lepage:1980fj,Ohrndorf:1981uz,Baier:1981pm}
\begin{align} \nonumber
\phi_{M}^{(1,8)}(x,\mu_F^2)&=\frac{6 f_{(1,8)}^M}{2\sqrt{N_C}} x(1-x)\large[1+\sum_{n=2,4,\cdots} a_n^{(1,8)}(\mu_F^2)C_n^{3/2}(2x-1)\large]\;,\\ \label{waves}
\phi_{M}^G(x,\mu_F^2)&=\frac{f_1^M}{2\sqrt{N_C}}\sqrt{\frac{C_F}{2n_f}} x(1-x)\sum_{n=2,4,\cdots} a_n^G(\mu_F^2) C_{n-1}^{5/2}(2x-1)\;,
\end{align}
where we follow the normalization convention described in~\cite{Harland-Lang:2013ncy}. The $f_{1,8}^M$ (with $M=\eta,\eta'$ in the present case) are given by (\ref{etafit}), with the $M$ dependence expressing the difference due to the mixing of the $\eta$, $\eta'$ states and decay constants. The evolution of the distribution amplitude is then dictated by the $\mu_F^2$ dependence of the coefficients $a_n$, see~\cite{Harland-Lang:2013ncy} for more details. In~\cite{Kroll:2013iwa} the contribution of the $gg$ Fock state to the transition form factors $F_{\eta(')\gamma}(Q^2)$ was investigated, and it was found that $a_2^G=19\pm 5$. However $a_2^G=0$ is still not necessarily excluded~\cite{DiDonato:2011kr}, in particular bearing in mind that the gluonic component only enters the transition form factors as an NLO correction, in contrast to  $\chi_c\to \eta(')\eta(')$ decays and central exclusive $\eta(')\eta(')$ production~\cite{Harland-Lang:2013ncy}.

To make contact with the physical $\eta$, $\eta'$ states we will be considering in this paper, we introduce the flavour--singlet and non--singlet quark basis states
\begin{align} \nonumber
|q\overline{q}_1\rangle &=\frac{1}{\sqrt{3}}|u\overline{u}+d\overline{d}+s\overline{s}\rangle\;,\\ \label{qfock}
|q\overline{q}_8\rangle &=\frac{1}{\sqrt{6}}|u\overline{u}+d\overline{d}-2s\overline{s}\rangle\;,
\end{align}
and the two--gluon state
\begin{equation}\label{gfock}
|gg\rangle\;.
\end{equation}
We then follow~\cite{Feldmann:1997vc} (see also~\cite{Schechter:1992iz,Kiselev:1992ms,Leutwyler:1997yr}) in taking a general two--angle mixing scheme for the $\eta$ and $\eta'$ mesons. That is, the mixing of the $\eta$, $\eta'$ decay constants is not assumed to follow the usual (one--angle) mixing of the states. This is most easily expressed in terms of the $\eta$ and $\eta'$ decay constants
\begin{align}\nonumber
f_8^\eta=f_8 \cos \theta_8\;,   \qquad\qquad f_1^\eta&=-f_1 \sin \theta_1 \;,\\ \label{etafit}
f_8^{\eta'}=f_8 \sin \theta_8\;,     \qquad\qquad f_1^{\eta'}&=f_1 \cos \theta_1 \;,
\end{align}
with the fit of~\cite{Feldmann:1998vh} giving
\begin{align}\nonumber
  f_8=1.26 f_\pi\;, \qquad & \qquad \theta_8   = -21.2^\circ\;,\\ \label{thetafit}
 f_1=1.17 f_\pi\;,  \qquad & \qquad \theta_1  = -9.2^\circ\;,
\end{align}
where we take  $f_\pi = 93$ MeV (for another approach see~\cite{Mathieu:2010ss}). We then take the distribution amplitudes (\ref{waves}) with the decay constants given as in (\ref{etafit}), for the corresponding Fock components (\ref{qfock}) and (\ref{gfock}). That is, the $\eta$ and $\eta'$ states are given schematically by
\begin{align} \nonumber
|\eta\rangle &= f_8\cos \theta_8 \left[\tilde{\phi}_{8,\eta}(x,\mu_F^2)|q\overline{q}_8\rangle \right]-f_1\sin \theta_1 \left[\tilde{\phi}_{1,\eta}(x,\mu_F^2)|q\overline{q}_1\rangle+\tilde{\phi}_{G,\eta}(x,\mu_F^2)|gg\rangle\right]\;,\\ \label{mixing}
|\eta'\rangle &= f_8\sin \theta_8 \left[\tilde{\phi}_{8,\eta'}(x,\mu_F^2)|q\overline{q}_8\rangle\right] +f_1\cos \theta_1 \left[\tilde{\phi}_{1,\eta'}(x,\mu_F^2)|q\overline{q}_1\rangle+\tilde{\phi}_{G,\eta'}(x,\mu_F^2)|gg\rangle\right]\;,
\end{align}
where to make things explicit the distribution amplitudes $\tilde{\phi}$ are defined as in (\ref{waves}), but with the decay constants divided out (i.e. $\tilde{\phi}_{8,\eta}(x,\mu_F^2)=\phi_{8,\eta}(x,\mu_F^2)/f_8^\eta$...), and these do not represent the conventional, normalized expressions for the $\eta',\eta$ Fock states, but simply indicate the decay constants and distribution amplitudes that should be associated with each $q\overline{q}$ and $gg$ state in this two--angle mixing scheme.

\section{Parton--level amplitudes}\label{sec:amp}

\begin{figure}
\begin{center}
\subfigure[]{\includegraphics[scale=0.8]{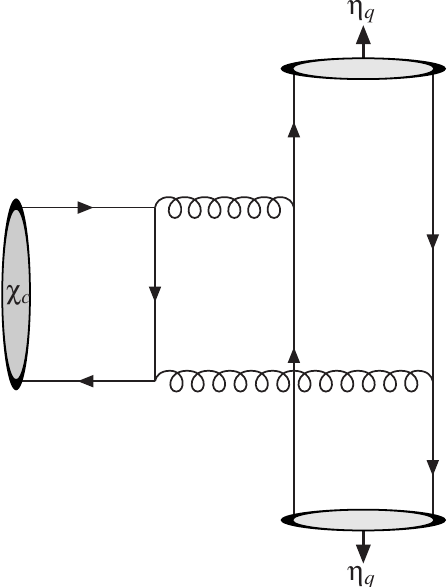}}\qquad\qquad\qquad
\subfigure[]{\includegraphics[scale=0.8]{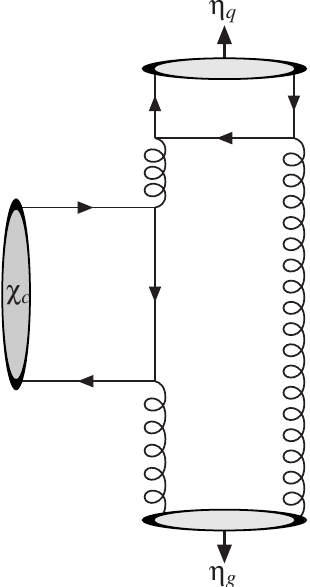}}\qquad\qquad\qquad 
\subfigure[]{\includegraphics[scale=0.8]{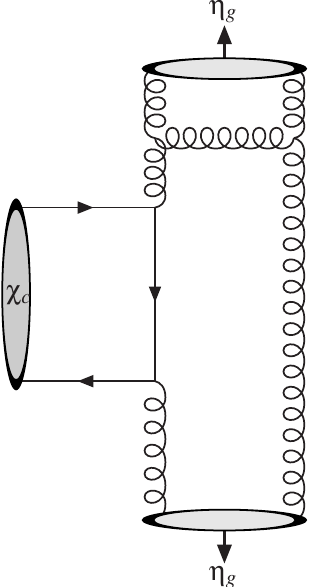}}
\caption{Representative Feynman diagrams for $\chi_c$ decays to $\eta(')\eta(')$ mesons, for (a) $q\overline{q}q\overline{q}$ (b) $q\overline{q}gg$ (c) $4g$ final states.}\label{feyn}
\end{center}
\end{figure}

Results for the relevant $\chi_{c(0,2)}\to q\overline{q}q\overline{q}$, $q\overline{q}gg$, $4g$ transitions were calculated in~\cite{Baier:1985wv}, see Fig.~\ref{feyn} for representative Feynman diagrams. To make contact with the experimentally measured branching ratios, we write
\begin{equation}
{\rm Br}(\chi \to M_1 M_2)=16\pi^2 C_F^2\frac{\alpha_s^2}{M_\chi^4}b\cdot K^2\;,
\end{equation}
with $b=8/9$ for the case of the $\chi(0+)$ and $b=4/3$ for the $\chi(2+)$. The above expression defines $K$, which is written in terms of the contributions
\be
K=K_{qq}+K_{qg}+K_{gg}\;,
\ee
from the quark and gluon final states. For the purely $q\overline{q}$ final state we have
\be
K_{qq}=-\int^1_0\int^1_0{\rm d}x_1\,{\rm d}x_2\,\frac{c(x_1,x_2)\phi^q_{M_1}(x_1)\phi^q_{M_2}(x_2)}{x_1(1-x_1)x_2(1-x_2)(1+(2x_1-1)(2x_2-1))}\ ,
\ee
where
\begin{align}
\chi_0:c(x_1,x_2)&=1+\frac{(1-x_1-x_2)^2}{1+(2x_1-1)(2x_2-1)}\;,\\
\chi_2:c(x_1,x_2)&=\frac{1}{2}-\frac{(1-x_1-x_2)^2}{1+(2x_1-1)(2x_2-1)}\;.
\end{align}
For the mixed $q\overline{q}$ and $gg$ final state we have\footnote{We note that the expression in~\cite{Baier:1985wv} for $K_{qg}$ in the case of the $\chi_c(0+)$ decay contains a misprint: in particular $K_{qg}=-(A_1+A_2)/2$ and not $K_{qg}=-(A_1-A_2)/2$. We are grateful to Andrey Grozin for confirming this.}
\be
K_{qg}=\pm \frac{1}{2}\sqrt{\frac{N_c}{N_c^2-1}}\left(I^G_1\sum_q I^q_2+I^G_2\sum_q I^q_1\right)\; ,
\ee
where the plus(minus) sign corresponds to the $\chi_2$($\chi_0$) decay. The values of $I^g$ and $I^q$ are
\be
I^{G,q}_M=\int_0^1 \frac{\phi^{G,q}_M(x)dx}{1-x}\ .
\ee
For the purely gluonic final state we have
\be
K_{gg}=\frac{N_c^2}{N_c^2-1}\int_0^1\int_0^1{\rm d}x_1 \,{\rm d}x_2\,\left(g_1+\frac{g_2}{N^2_c}\right)\phi^g_{M_1}(x_1)\phi^g_{M_2}(x_2)\;,
\ee
with
\begin{align}
g_1&=X_1X_2\beta_1\beta_2\beta_{12}[11-6(\xi_1+\xi_2)+\xi_1^2+\xi^2_2-8\xi_1\xi_2+2\xi_1\xi_2(\xi_1+\xi_2)+3\xi^2_1\xi^2_2]\;,\\
g_2&=-X_1X_2\beta_{12}(4-\xi_1-\xi_2-2\xi_1\xi_2)\;,
\end{align}
for the case of the $\chi(0+)$ decay and
\be
g_1=X_1X_2\beta^2_{12}(2-\xi_1-\xi_2)\;\;\;\;\;\;\; g_2=-X_1X_2\beta^2_{12}(1-\xi_1)(1-\xi_2)\;, 
\ee
for the case of the $\chi(2+)$. Here we have defined $X_i=2x_i-1$, $\xi_i=X^2_i$, $\beta_i=1/(1-\xi_i)$ and $\beta_{12}=1/(1-\xi_1\xi_2)$.

We note that our results differ from those of~\cite{Baier:1985wv} by overall factors due our differing normalization convention for the meson distribution amplitudes, although the combined amplitude as in (\ref{mab}) is of course consistent. In addition, we have associated an additional factor of $(-1)$ to the $K_{qq}$ term due to the permutation of fermionic operators corresponding to this amplitude, see~\cite{Harland-Lang:2013ncy} for more details. 

\section{Numerical results}\label{sec:num}

We apply the results of the previous sections to calculate the branching ratios of the $\chi_{c0}$ and $\chi_{c2}$ to $\eta(')\eta(')$ mesons, checking also the $\chi_c\to \pi\pi$ case to confirm that the decay to this purely $q\overline{q}$ final state is sufficiently well described. For the singlet and octet quark distribution amplitudes we will make use of the `CZ' form (\ref{CZ}), while for the gluonic component we will consider a range of values for the first Geigenbaur coefficients $a_2^G$ (as in~\cite{Harland-Lang:2013ncy} we can safely omit the small corrections from higher $n$ terms). As the scale $\mu_F \sim m_c$ is of the same order as the typical input scale $\mu_0 \sim 1$ GeV for the meson distribution amplitudes, we for simplicity do not include any wave function evolution, i.e. we fix $\mu_F=\mu_0$. We take $\alpha_s=0.335$, which in combination with a reasonable choice for the derivative of the $\chi_c$ wave function at the origin provides a fairly good NLO description of the total and radiative widths of the $\chi_c(0+)$ and the $\chi_c(2+)$ mesons (see e.g.~\cite{Diakonov:2012vb}). In order to account for the relatively large masses of the $\eta(')$ mesons we multiply the result by the factor $(2p/M_\chi)^{2J+1}$, where $p$ is the magnitude of the meson 3--momentum in the $\chi_c$ rest frame, and $J=0,2$ is the spin of the $\chi_{c(0,2)}$ meson. This accounts for the available phase space and orbital angular momentum of the two final--state meson.

\begin{table}
\begin{center}
\begin{tabular}{|c|c|c|c|}\hline
$a_2^G$ & $\eta\eta$ & $\eta'\eta'$ & $\eta\eta'$  \\
\hline
 -32&0.48 & 0.093 & 0.22\\
 -16&0.51 & 0.18 & 0.087\\
0&0.52 & 0.54 & 0.050\\
16&0.51 & 0.38 & 0.063\\
32&0.49 & 0.0047 & 0.14\\
\hline
Experiment~\cite{Olive:2016xmw} & 0.295 & 0.196 &  $<0.02$\\
\hline
\end{tabular}
\end{center}
\caption{\sf Predicted $\chi_{c0}\to \eta(')\eta(')$ branching ratios (in \%) for different values of the coefficient $a_2^G$ from (\ref{waves}), corresponding to the normalization of the $gg$ component for the flavour--singlet $\eta_1$ state.}\label{chi0}
\end{table}

\begin{table}
\begin{center}
\begin{tabular}{|c|c|c|c|}\hline
$a_2^G$ & $\eta\eta$ & $\eta'\eta'$ & $\eta\eta'$ \\
\hline
-32 & 0.088 & 0.13 & 0.0023 \\
-16 & 0.087 & 0.093 & 0.0039\\
0 & 0.086 & 0.055 & 0.0066\\
16 & 0.084 & 0.023 & 0.011 \\
32 & 0.082 & 0.0032 & 0.017\\
\hline
Experiment~\cite{Olive:2016xmw} & 0.057 & $<0.01$ &  $<0.006$  \\
\hline
\end{tabular}
\end{center}
\caption{\sf Predicted $\chi_{c2}\to \eta(')\eta(')$ branching ratios (in \%) for different values of the coefficient $a_2^G$ from (\ref{waves}), corresponding to the normalization of the $gg$ component for the flavour--singlet $\eta_1$ state.}\label{chi2}
\end{table}

Firstly, to check that the formalism of the preceding sections can be used reliably for a purely $q\overline{q}$ final state, we can calculate the branching ratio for the $\chi_{c}\to \pi\pi$ decays, including both the neutral and charged channels. Applying the `CZ' distribution amplitude of (\ref{CZ}) we find
\begin{align}
{\rm Br}(\chi_{c0}\to \pi\pi)&=0.82\%\;,\\
{\rm Br}(\chi_{c2}\to \pi\pi)&=0.16\%\;,
\end{align}
to be compared with the experimental values~\cite{Olive:2016xmw} of ${\rm Br}(\chi_{c0}\to \pi\pi)=(0.833\pm0.035)\%$ and  ${\rm Br}(\chi_{c2}\to \pi\pi)=(0.233\pm0.012)\%$. The description is excellent in the case of the $\chi_{c0}$, but is somewhat low in the $\chi_{c2}$ case. However, given the possibility that higher order QCD (as well as higher twist) corrections may give additional $\sim 30-50\%$ spin--dependent corrections to these simple lowest order predictions (this is seen in for example the case of the total $\chi_c$ widths~\cite{Barbieri:1980yp,Barbieri:1981xz,Lansberg:2009xh}), these results can give us confidence in our predictions for the $\eta(')$ final states at this level of accuracy.

\begin{table}
\begin{center}
\begin{tabular}{|c|c|c|c|}\hline
$a_2^G$ & $\eta\eta$ & $\eta'\eta'$ & $\eta\eta'$  \\
\hline
-32 & 0.29 & 0.24 & 0.057\\
-16 & 0.32 & 0.056 & 0.0035\\
0 & 0.33 & 0.28 & 3.6 $\times 10^{-4}$ \\
16 & 0.33 & 0.16 & 1.8 $\times 10^{-5}$  \\
32 & 0.31 & 0.024 & 0.017 \\
\hline
Experiment~\cite{Olive:2016xmw} & 0.295$\pm 0.019$ & 0.196$\pm 0.021$ &  $<0.023$  \\
\hline
\end{tabular}
\end{center}
\caption{\sf As in Table~\ref{chi0}, but including strange quark suppression.}\label{chi0s}
\end{table}

In Table~\ref{chi0} we show the predicted $\chi_{c0}\to \eta(')\eta(')$ branching ratios for a range of different values of the coefficient $a_2^G$ from (\ref{waves}), corresponding to the normalization of the $gg$ component for the flavour--singlet $\eta_1$ state. We can see that as $|a_2^G|$ is increased from zero the $\eta'\eta'$ branching ratio in fact decreases in size. This effect is driven by the destructive interference between the purely $q\overline{q}$ and purely $gg$ contributions, $K_{qq}$ and $K_{gg}$, respectively, as defined in Section~\ref{sec:calc}; for $|a_2^G|\sim 30$ this interference is almost complete. In addition, there is some dependence on the sign of $a_2^G$ driven by the mixed contribution $K_{qg}$. These effects allow the qualitative trend in the data, in particular the dominance of the $\eta\eta$ mode, to be described for reasonable choices of $a_2^G$. However, the branching ratios in all three channels are in general larger than the observed values, in particular for the mixed $\eta\eta'$ decay. The $\chi_{c2}$ case is shown in Table~\ref{chi2} and a similar trend is found, with the disagreement being even more severe.

\begin{table}
\begin{center}
\begin{tabular}{|c|c|c|c|}\hline
$a_2$ & $\eta\eta$ & $\eta'\eta'$ & $\eta\eta'$  \\
\hline
-32 & 0.058 & 0.077 & 0.0025  \\
-16 & 0.057 & 0.053 & 9.1 $\times 10^{-4}$ \\
0 & 0.055 & 0.029 & 4.7 $\times 10^{-5}$   \\
16 & 0.053 & 0.0094 & 4.1 $\times 10^{-4}$   \\
32 & 0.050 & 9.1 $\times 10^{-5}$ & 0.0026 \\
\hline
Experiment~\cite{Olive:2016xmw} & 0.057$\pm 0.005$ & $<0.01$ &  $<0.006$  \\
\hline
\end{tabular}
\end{center}
\caption{\sf As in Table~\ref{chi2}, but including strange quark suppression.}\label{chi2s}
\end{table}

The situation is however greatly improved if we consider the potential impact of the non--zero strange quark mass. In particular, as the mass of the $\chi_c$ is not so large, the strange quark mass may not be negligible in comparison to the average virtuality of the four quark propagators in Fig.~\ref{feyn} (a). If we take an average virtuality of $\langle q^2_s\rangle \sim M_\chi^2/4 \sim 0.75\,{\rm GeV}^2$, and take a strange quark mass of $m_s=0.25$ GeV (somewhere between the constituent and current quark values), then the contribution form each $s\overline{s}$ pair will be suppressed by a factor $F_s=1-2m^2_s/\langle q^2_s\rangle \sim 0.8$. In Tables~\ref{chi0s} and~\ref{chi2s} we show the same results as before, but including this strange quark suppression factor. We can see that the impact of this is quite significant, and that a good description of all experimental observations and limits on the $\chi_c$ branching ratios can be achieved for $a_2^G \sim 16$. Interestingly, this is in seen to be in good agreement with the result of~\cite{Kroll:2013iwa}, which found $a_2^G=19\pm 5$ from an analysis of the meson transition form factors $F_{\eta(')\gamma}(Q^2)$.

\section{Conclusion}\label{sec:conc}

In this paper we have examined the role of interference and strange quark mass effects in the decay of $\chi_{c(0,2)}$ mesons to $\eta(')$ pairs. This was motivated by the apparent tension between the general requirement for a sizeable $gg$ component of the $\eta(')$ meson and the fact that the branching ratios for the $\chi_c$ decays to $\eta'\eta'$ mesons, which occurs via a purely gluonic intermediate state, is in fact observed to lie below the case of the $\eta\eta$ decays. We have in particular demonstrated that the amplitudes for the $\chi_c$ transitions to purely quark and purely gluon final states interfere destructively, naturally leading to a suppression in the $\eta'\eta'$ mode. We have then shown explicitly that for a reasonable choice of the $gg$ component of the $\eta(')$ (and through mixing, the $\eta$) the experimentally observed branchings can be reproduced, resolving this apparent tension.

Our numerical results have in addition included the impact of strange quark mass suppression in the $\chi_c$ decays. That is, for the relatively low $\chi_c$ masses, the virtuality carried by strange quark propagators in the $\chi_c \to \eta(')\eta(')$ decay may not be significantly greater than the strange quark mass. Indeed, including a simple model for this effect we find the impact on the predicted branching ratios can be quite large. Combining these effects, we have shown that for a sensible choice of the first Geigenbaur coefficient of the gluon component of the flavour singlet $\eta_1$ wave function, $a_2^G\sim 16$, we can reproduce all of the experimentally observed $\chi_c\to \eta(')\eta(')$ branching ratios, as well as results below the corresponding limits in the absence of current observations. This value is consistent with the separate study of~\cite{Kroll:2013iwa} where the gluonic component of the $\eta(')$ meson is extracted from an analysis of the meson transition form factors $F_{\eta(')\gamma}(Q^2)$.

It is worth emphasising that the results of this paper, while encouraging, are only calculated at leading order: it is well known for example that the higher order $\alpha_s$ corrections for the $\chi_c$ total decay widths are rather large and in fact have different signs for the $\chi_c(0+)$ and the $\chi_c(2+)$ cases~\cite{Barbieri:1980yp,Barbieri:1981xz,Lansberg:2009xh}. The inclusion of these corrections, including colour octet contributions which enter at this order~\cite{Bolz:1997ez}, may therefore have a non--negligible impact on the precise quantitative results. Moreover, a more complete treatment of strange quark mass effects may also have some impact.

 Therefore, while the results of this paper demonstrate for the first time the role that interference effects play in allowing a realistic $gg$ component for the $\eta'$ when considering these $\chi_c$ decays, further study is certainly needed to clarify these issues further. A simple test would be to compare against future measurements (rather than the existing upper limits) of the $\chi_{c0}\to \eta\eta'$ and $\chi_{c2}\to\eta'\eta'$ and $\eta\eta'$ decays. It would in addition be interesting to increase the scale at which the gluon component of the $\eta'$ is probed. For example, the decay of $\chi_b$ mesons will be similarly sensitive to the effects described in this paper, but safely in the region where strange quark mass effects will be negligible, providing a clearer test. Alternatively, by observing central exclusive $\eta(')\eta(')$ meson pair production at the LHC for reasonable $\eta(')$ meson transverse momentum $p_\perp$, a direct and clear probe of the gluon component can be provided~\cite{Harland-Lang:2013ncy}.

\section*{Acknowledgements}

We are grateful to  Andrey  Grozin for useful discussions. MGR thanks the IPPP at the University of Durham for hospitality. The work of MGR and AGS was supported by the RSCF grant 14-22-00281. VAK thanks the Leverhulme Trust for an Emeritus Fellowship. LHL thanks the Science and Technology Facilities Council (STFC) for support via the grant award ST/L000377/1.

\bibliography{references}{}
\bibliographystyle{h-physrev}

\end{document}